\documentclass[prl,aps,secnumarabic,showpacs,twocolumn]{revtex4}
\usepackage{bm,latexsym,amsmath,amsfonts,graphicx,subfigure,color}
\usepackage{hyperref,url,hhline,multirow}

\allowdisplaybreaks[4]

\begin{document}


\title{Residual Symmetries for Neutrino Mixing with a Large $\theta_{13}$
and Nearly Maximal $\delta_D$}
\author{
Shao-Feng Ge$^{1,}$\footnote{Electronic address: gesf02@gmail.com}, 
Duane A. Dicus$^{2,}$\footnote{Electronic address: dicus@physics.utexas.edu}, 
and Wayne W. Repko$^{3,}$\footnote{Electronic address: repko@pa.msu.edu}}
\affiliation{
$^1$Institute of Modern Physics and Center for High Energy Physics, Tsinghua University, Beijing 100084, China  \\
$^2$Physics Department, University of Texas, Austin, TX 78712  \\
$^3$Department of Physics and Astronomy, Michigan State University, East Lansing MI 48824}

\date{December 8, 2011}

\begin{abstract}
The residual $\mathbb Z^s_2(k)$ and $\overline{\mathbb Z}^s_2(k)$ symmetries induce a direct and
unique phenomenological relation with $\theta_x (\equiv \theta_{13})$ expressed in terms 
of the other two mixing angles, $\theta_s (\equiv \theta_{12})$ and $\theta_a (\equiv \theta_{23})$,
and the Dirac {\tt CP} phase $\delta_D$. 
$\mathbb Z^s_2(k)$ predicts a $\theta_x$ probability distribution 
centered around $3^\circ \sim 6^\circ$ with an uncertainty of $2^\circ$ to $4^\circ$ while
those from $\overline{\mathbb Z}^s_2(k)$ are approximately a factor of two larger.
Either result fits the T2K, MINOS and Double Chooz measurements. Alternately a prediction for 
the Dirac {\tt CP} phase $\delta_D$ 
results in a peak at $\pm 74^\circ$ ($\pm 106^\circ$) for $\mathbb Z^s_2(k)$
or $\pm 123^\circ$ ($\pm 57^\circ$) for $\overline{\mathbb Z}^s_2(k)$ which is consistent with the 
latest global fit. 
We also give a distribution for the  leptonic Jarslkog invariant $J_\nu$  
which can provide further tests from measurements at T2K and NO$\nu$A.
\end{abstract}
\pacs{14.60.Pq \hfill Accepted for publication in PRL [arXiv:1108.0964]}
\maketitle

{\it Introduction} --
The T2K \cite{T2K} and MINOS \cite{MINOS} experiments  
indicate a relatively large reactor angle $\theta_x$ for neutrino mixing. At the 90\% C.L., T2K gives 
$0.03\, (0.04) < \sin^2 2 \theta_x < 0.28\, (0.34)$, with zero Dirac {\tt CP} phase, $\delta_D$, 
for normal (inverted) hierarchy while MINOS gives $0.01\,(0.026) < \sin^2 2 \theta_x < 0.088\,(0.150)$
and Double Chooz \cite{doublechooz} with $\sin^2 2 \theta_{13} = 0.085 \pm 0.051$ at 68\% C.L..

Many varied theoretical efforts have been made to understand this large $\theta_x$.
Discrete groups such as $S_3$ \cite{Zhou:2011nu}, $A_4$ \cite{A4a,King:2011zj},
$S_4$ \cite{King:2011zj,S4a,Meloni:2011fx},
and the binary tetrahedral group $T'$ \cite{Tprime}
have been quite popular while new possibilities are explored in \cite{new}.
Other efforts concentrate on perturbations from some featured zeroth-order mixing
such as democratic \cite{Xing:2011at,Chao:2011sp}, 
bimaximal \cite{Meloni:2011fx,Marzocca:2011dh,Chao:2011sp}, 
tribimaximal \cite{King:2011zj,Marzocca:2011dh,tribi},
and tetra-maximal \cite{Zhang:2011aw} patterns. 
More discussions can be found in \cite{more}.

In these papers symmetries or other model assignments are employed. We will show that
phenomenological consequences of residual symmetries $\mathbb Z^s_2(k)$ 
and $\overline{\mathbb Z}^s_2(k)$ can be readily established predicting not only $\theta_x$ to be large, 
fitting the T2K, MINOS and Double Chooz data, but also $\delta_D$ nearly maximal in good agreement 
with the latest global fits.
This provides the first strong and direct evidence for residual symmetries.

{\it Residual Symmetries} --
The symmetry that directly determines the lepton mixing pattern need not be the same as 
the full symmetry of the fundamental lagrangian. As the left-handed charged lepton and 
neutrino reside in a same $SU(2)_L$ doublet, they are governed by a common symmetry which
must be broken. Otherwise they would share a same diagonalization matrix \cite{Lam:2005va,Lam:horizontal}, 
leading to trivial leptonic mixing. It is the residual symmetry that determines the 
mixing matrices, if indeed the mixing is believed to be determined by symmetry.

It is convenient to work in the diagonal basis of charged leptons \cite{Dicus:2010iq}. 
To completely determine
the mixing matrix, a product of two $\mathbb Z_2$ symmetries is enough \cite{Lam:horizontal,arXiv:0906.2689}. 
One is the well-known $\mu$--$\tau$ symmetry \cite{mutau} and the other
is $\mathbb Z^s_2$ \cite{Lam:horizontal} which can be extended to accommodate
a general solar angle \cite{Ge:2010js} generated by,
\begin{equation}
  G_1(k)
=
  \frac{1}{2+k^2}
  \left\lgroup
  \begin{matrix}
    2-k^2 & 2k & 2k  \\
    2k & k^2 & -2  \\
    2k & -2 & k^2
  \end{matrix}
  \right\rgroup \,.
\label{G1}
\end{equation}
There is another residual $\overline{\mathbb Z}^s_2(k)$ represented by $G_2 \equiv G_1 G_3$,
where $G_3$ is the 
matrix for $\mu$--$\tau$ symmetry \cite{Lam:horizontal},
\begin{equation}
  G_2(k)
=
  \frac{1}{2+k^2}
  \left\lgroup
  \begin{matrix}
    2-k^2 & 2k & 2k  \\
    2k & -2 & k^2  \\
    2k & k^2 & -2
  \end{matrix}
  \right\rgroup \,.
\label{G2}
\end{equation}
Since $\mu$--$\tau$ symmetry is just a first order approximation, indicated by the experimental data
\cite{T2K, MINOS, doublechooz} and other considerations \cite{mutau}, it has to be broken.
The remaining symmetry would be $\mathbb Z^s_2(k)$ or $\overline{\mathbb Z}^s_2(k)$ but not both 
as they are not independent. However, their phenomenological consequences need not be 
the same as we show below. Note that, since the diagonal mass matrix of charged leptons is not
degenerate, $G_1$ and $G_2$ only apply to the neutrino sector after the full symmetry is broken down
to residual symmetries.

{\it Correlation Between Mixing Angles} --
With a single $\mathbb Z^s_2(k)$, a correlation between the three mixing angles and the
Dirac {\tt CP} phase can be derived.  In particular,
\begin{subequations}
\begin{eqnarray}
  \cos \delta_D
=
  \frac {(s^2_s - c^2_s s^2_x)(c^2_a - s^2_a)}
        {4 c_a s_a c_s s_s s_x} \,,
\label{eq:G-relation-a}
\\
  \cos \delta_D
=
  \frac {(s^2_s s^2_x - c^2_s)(c^2_a - s^2_a)}
        {4 c_a s_a c_s s_s s_x} \,,
\label{eq:G-relation-b}
\end{eqnarray}
\label{eq:G-relation}
\end{subequations}
for $\mathbb Z^s_2(k)$ \cite{Ge:2011ih} and $\overline{\mathbb Z}^s_2(k)$ respectively.
Note that only physical quantities are involved which gives the possibility of robust physical 
predictions. By implementing the measured values of the three mixing angles, a prediction of 
$\delta_D$ can be made. Or, (\ref{eq:G-relation}) can be solved for $\theta_x$,
\begin{equation}
\hspace{-3mm}
  \sin \theta_x
=
  p
  \left[
  \pm \sqrt{c^2_D + \cot^2 2 \theta_a}
  - c_D
  \right] \tan 2 \theta_a (\tan \theta_s)^p,
\label{eq:thetax}
\end{equation}
with $c_D \equiv \cos \delta_D$ while $p = \pm\,1$ for $\mathbb Z^s_2(k)$ or 
$\overline{\mathbb Z}^s_2(k)$. 
Solutions for the $\pm$ sign within the parenthesis are equivalent
through a redefinition $(\theta_x, \delta_D) \rightarrow (-\theta_x, \delta_D + \pi)$
leaving no effect on the measured physical quantity $\sin^2 \theta_x$. This is also 
true for the overall $p$. The difference comes from the exponent $p$ 
leading to a $(\tan \theta_s)^2 \approx 1/2$ factor between the $\mathbb Z^s_2(k)$ and
$\overline{\mathbb Z}^s_2(k)$ predictions.

The main feature of (\ref{eq:thetax}) can be seen by expanding it to the leading 
order. As the reactor angle $\theta_x \equiv \delta_x$ is small and the atmospheric angle 
$\theta_a \equiv 45^\circ + \delta_a$ is nearly maximal, (\ref{eq:G-relation})
reduces to,
\begin{equation}
  \frac {\delta_x}{\delta_a}
=
- p \frac {(\tan \theta_s)^p}{\cos \delta_D} \,.
\label{eq:ratio}
\end{equation} 
 
Eqs.(\ref{eq:G-relation})--(\ref{eq:ratio}) are general and direct.  
To demonstrate this, three examples are provided. Eq.(\ref{eq:ratio})
was first obtained in a minimal seesaw model \cite{Ge:2010js} with $\mu$--$\tau$
and {\tt CP} softly broken and $\mathbb Z^s_2(k)$ retained exactly, befitting the situation 
discussed here. A special case with $k = 2$, which constrains the mixing matrix
to be trimaximal, is studied in 
\cite{King:2011zj}. Even an 
"unphysical" bimaximal solution \cite{Lam:2011zm} can be covered as a marginal example. 
Note that the first two examples are obtained in model-dependent and perturbative ways while
the last one comes from a pure symmetry analysis. 

The ratio (\ref{eq:ratio}) 
of the deviation of the reactor angle from zero and that of the atmospheric angle from maximal
is given by the solar angle and  the CP phase.
Its absolute value is a  minimum when $\delta_D$
equals $0$ or $\pi$, $|\delta_x| \geq (\tan \theta_s)^p |\delta_a|$. 
Alternately (\ref{eq:G-relation}) can be solved exactly with $\cos \delta_D\,=\,\pm\,1$ to give
an absolute lower bound
$\sin \theta_x \geq (\tan \theta_s)^p |c_a - s_a|/|c_a + s_a|$.
An upper bound can also be obtained but since $c_a \approx s_a$ it is larger than $1$.

{\it Numerical Predictions} --
A nonzero $\theta_x$ has been consistent with global fits for several years. 
The first hint appears in \cite{Balantekin:2008zm} at only $0.9 \sigma$ C.L.. 
It persists in all subsequent global fits 
\cite{Fogli:2008jx,Schwetz:2008er,Maltoni:2008ka,GonzalezGarcia:2010er,Schwetz:2011qt}
and increases steadily to about $3 \sigma$ \cite{Fogli:2011qn,Schwetz:2011zk} as
summarized in Table~\ref{tab:data}.
\begin{table}[h]
\begin{center}
{\small
\begin{tabular}{|c||c|c|c|}
\hline
& & & \\[-3mm]
& $\sin^2\theta_s\,(\theta_s)$
& $\sin^2\theta_a\,(\theta_a)$ 
& $\sin^2\theta_x\,(\theta_x)$
\\
& & & \\[-3mm]
\hhline{|=::=|=|=|}
& & & \\[-4mm]
Best Fit
& $0.306$~($33.6^\circ$)
& $0.42$~($40.4^\circ$) 
& $0.021$~($8.3^\circ$)
\\[1mm]
\hline
& & & \\[-3mm]
  \multirow{2}{*}{$1\sigma$ Range}
& $0.291$-$0.324$ 
& $0.39$-$0.50$ 
& $0.013$-$0.028$
\\[0.6mm]
& ($32.7$-$34.7^\circ$) 
& ($38.7$-$45.0^\circ$) 
&  ($6.6$-$9.6^\circ$)
\\[1mm]
\hline
\end{tabular}
} 
\caption{The global fit \cite{Fogli:2011qn} for the neutrino mixing angles.}
\label{tab:data}
\end{center}
\end{table}

The fits can be classified into two catagories depending on the result for the atmospheric 
angle $\theta_a$ 
which is persistently maximal in 
\cite{Schwetz:2008er,Maltoni:2008ka,Schwetz:2011qt,Schwetz:2011zk} while an apparent deviation from $45^{\circ}$ is 
claimed possible in \cite{Fogli:2008jx,Fogli:2011qn,GonzalezGarcia:2010er}
due to a subleading effect \cite{Fogli:2005cq}.
\begin{figure}[h]
\centering
\includegraphics[width=8cm,height=5cm]{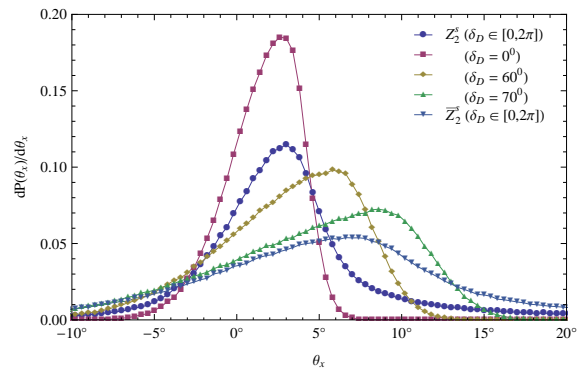}
\caption{(Color Online) Predicted distributions of $\theta_x$.}
\label{fig:thetax}
\end{figure}

From (\ref{eq:thetax}) the distribution of $\theta_x$ can be derived by using asymmetric Gaussian distributions ${\mathbb P}$ as,
\begin{equation}
  \frac {d P(\theta_x)}{d \theta_x}
=
  \int f^p_x {\mathbb P}(s^2_a) {\mathbb P}(s^2_s) d s^2_a d s^2_s \frac {d \delta_D} {2 \pi},
\label{eq:dP}
\end{equation}
where $f^p_x \equiv \frac 1 2 \delta (\theta_x - \arcsin \bar s_x)$ are $\delta$-functions
that pick out the predicted value $\bar s_x \equiv \mbox{RHS of}~(\ref{eq:thetax})$ for 
$\theta_x$, given a concrete input of $p, s^2_a$, $s^2_s$, and $\delta_D$. There is also a 
mirror contribution for negative $\theta_x$ that is not shown -- hence the $\frac 1 2$ prefactor. 
We take $\delta_D$ to be evenly distributed
in $[0, 2 \pi)$, as in (\ref{eq:dP}), or replace it with a specific value. The integration (\ref{eq:dP}) can 
be simulated with scattering points
or the delta function can be converted to one for $\theta_s$ and the other integrals done numerically.
The results are shown in Fig.~\ref{fig:thetax} for both 
$\mathbb Z^s_2(k)$ and $\overline{\mathbb Z}^s_2(k)$. We will first discuss the results from 
$\mathbb Z^s_2(k)$.

After averaging over $\delta_D$,
the probability distribution peaks around $3^\circ$ with an asymmetric width from 
$2^\circ$ to $4^\circ$. This is in sharp contrast with the distribution given by the 
previous global fit \cite{Ge:2011ih}, which peaks at $0^\circ$. 
From (\ref{eq:ratio}) we can see that $\delta_x$ is 
proportional to $\delta_a$. With $\theta_a$ significantly deviating from the maximal 
value, the predicted $\delta_x$ must increase accordingly. In the global fit
adopted in \cite{Ge:2011ih}, the central value of $\theta_a$ is about $43^\circ$ and the
maximal value is well within the $1 \sigma$ range. Hence, there is no apparent nonzero peak
in the predicted distribution of $\theta_x$. For the latest global fit \cite{Fogli:2011qn} of $\theta_a$ 
the central value is about $40.4^\circ$, 
while the maximal value is at the edge of the $1 \sigma$ region.
This significant change in  $\theta_a$ leads to a clear nonzero prediction of $\theta_x$. 
As $\theta_s$ only contributes as an overall factor in (\ref{eq:ratio}), its deviation will not
change the conclusion for $\theta_x$. For example, a different treatment of the reactor data leads to
$3.7\%$ difference \cite{Schwetz:2011qt} in $\tan \theta_s$.
The 
best fit values of 
$\sin^2 \theta_x (\theta_x)$  vary from $0.02 \,(8.1^\circ)$ 
to $0.04\, (11.5^\circ)$ which are still covered by our predictions. This is also treated in
\cite{Fogli:2011qn} but with a much smaller variation, approximately $20\%$ in $\sin^2 \theta_x$.

The measured $\theta_x$ \cite{T2K,MINOS,doublechooz} is not independent of the Dirac {\tt CP} phase $\delta_D$. 
But this does not affect the matching between the experimental result and the theoretical predictions.
Figure \ref{fig:thetax} shows that averaging over possible $\delta_D$ values gives a
best fit value and deviation which resemble those with vanishing $\delta_D$. As the experimental
fit of $\theta_x$ depends only slightly on $\delta_D$ while the theoretical prediction is sensitive
to it, as shown in (\ref{eq:ratio}), varying $\delta_D$ can effectively improve the matching. 
For example, using $\delta_D = 60^\circ$ moves the peak to the MINOS and Double Chooz central values 
while $\delta_D = 70^\circ$ for that of T2K.


Since $|\delta_x| \ge |\delta_a| \tan \theta_s$, the prediction for $\theta_x$ with vanishing $\delta_D$ 
is the most conservative in the sense that it gives the smallest prediction for $\theta_x$. 
The prediction with $\delta_D$ uniformly distributed also peaks at $3^\circ$ but has an extended tail to 
higher $\theta_x$. This is because, while $\delta_D$ is uniformly distributed, $\cos \delta_D$ is not. Its 
distribution varies as $(\sin \delta_D)^{-1}$ which is relatively suppressed for small $\cos \delta_D$. 
Thus, the most conservative region is the most probable one.
For example, the probability for $|\cos \delta_D| \leq 0.1\, (0.2, 0.3,0.4)$ is
just 6\% (13\%, 19\%, 26\%) corresponding to $\delta_D = 84^\circ\, (78^\circ, 73^\circ, 66^\circ)$
respectively. Most of the significant region lies between $\delta_D = 0^\circ$ and approximately
$\delta_D = 60^\circ$. 
Within this region, the $\theta_x$
peak varies from approximately $3^\circ$ to around $6^\circ$ and the width changes from roughly
$2^\circ \sim 4^\circ$ to almost $4^\circ \sim 8^\circ$. This is the region covered by MINOS
result $2.9^\circ (4.6^\circ) < \theta_x < 8.6^\circ (11.4^\circ)$ and
$5.3^\circ < \theta_x < 10.8^\circ$ for Double Chooz at $1\sigma$ level while T2K has
$5.0^\circ (5.8^\circ) < \theta_x < 16.0^\circ (17.8^\circ)$ at 90\% C.L..

The above discussion also applies to the case of $\overline{\mathbb Z}^s_2(k)$. The only difference 
is the factor of about $2$ coming from the exponent $p$ in (\ref{eq:thetax}). As $\theta_x < 10^\circ$ 
can be treated as small perturbation, this will induce  approximate factors of $2$ in the peak 
location and $1/2$ in its height relative to the predictions from ${\mathbb Z}^s_2(k)$. 
The result is still in good agreement with the data and the global fits.

\begin{figure}
\centering
\subfigure[\, Dirac {\tt CP} Phase $\delta_D$]
          {\label{fig:DJ-D} \includegraphics[width=8cm,height=5cm]{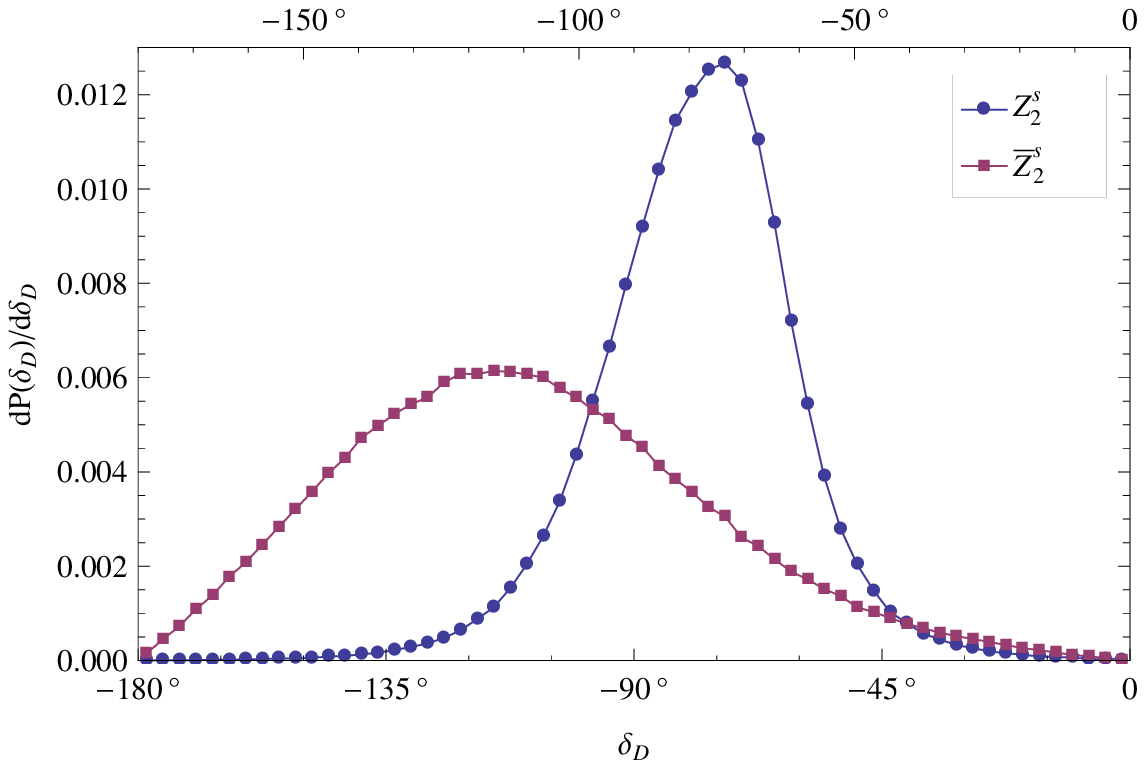}}
\qquad
\subfigure[\, Jarlskog Invariant $J_\nu$]
          {\label{fig:DJ-J} \includegraphics[width=8cm,height=4.7cm]{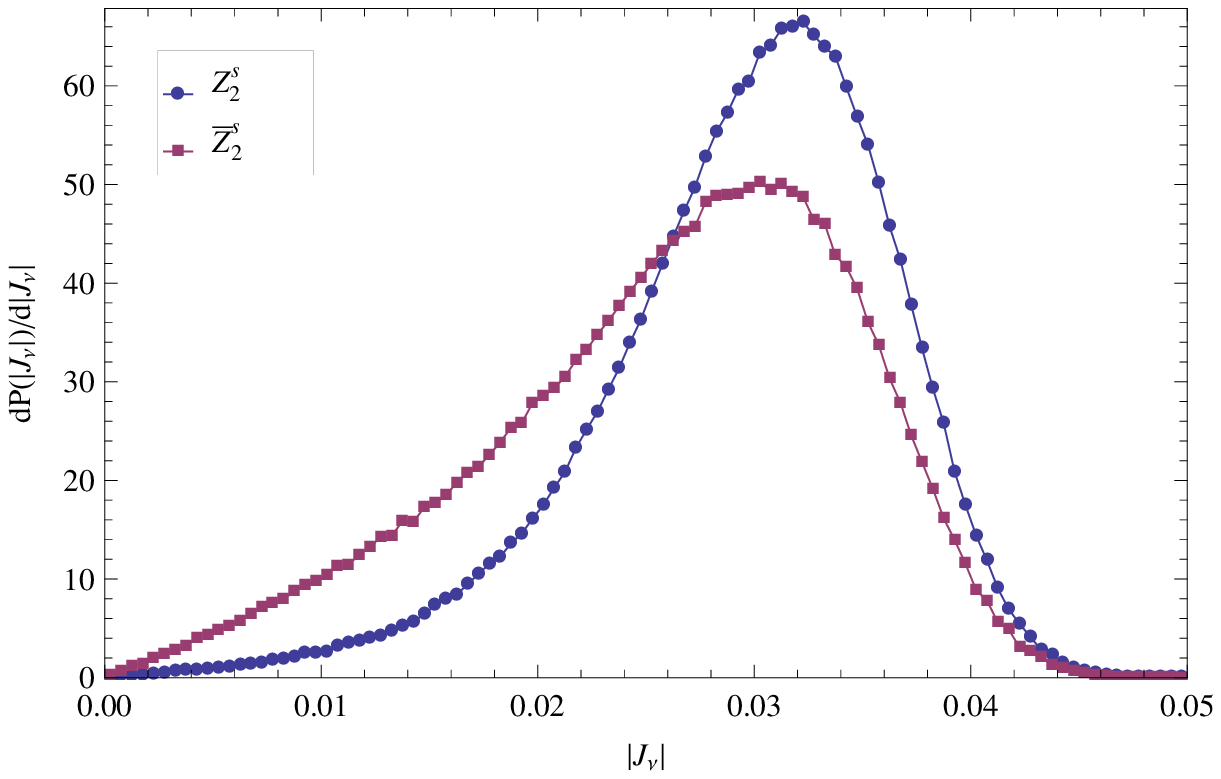}}
\caption{(Color Online) Predicted distributions of (a) the Dirac {\tt CP} phase $\delta_D$
         and (b) Jarlskog Invariant $J_\nu$.}
\label{fig:DJ}
\end{figure}

This consistency between the data and our prediction of a large 
$\theta_x$ provides the first nontrivial indication of the viability of residual symmetries 
$\mathbb Z^s_2(k)$ or $\overline{\mathbb Z}^s_2(k)$. The correlation between
the mixing angles (\ref{eq:G-relation}) is independent of the group parameter $k$, and is 
obtained in a direct way, making the result quite robust.

The change in the global fit also alters our prediction 
of the Dirac {\tt CP} phase $\delta_D$ \cite{Ge:2011ih}. As shown in Fig.~\ref{fig:DJ-D} the most
probable value of $\delta_D$ is no longer maximal. This is also caused by the shifted central 
value of $\theta_a$. As $\delta_a$ deviates further from $\frac{\pi}{4}$, maximal $\delta_D$ 
becomes less probable as indicated by (\ref{eq:G-relation}). Instead it peaks around $\pm 74^\circ$
for $\mathbb Z^s_2$. 
Notice that a mirror solution in (\ref{eq:thetax}) can be obtained through 
$(\theta_x, \delta_D) \rightarrow (- \theta_x, \delta_D + \pi)$ generating another peak
around $\pm 106^\circ$. These are in perfect consistency with the indication of 
$-74^\circ (-110^\circ)$ for inverted (normal) hierarchy \cite{Schwetz:2011zk}.
Although no concrete number is provided, a nonzero {\tt CP} phase also appears in \cite{Fogli:2011qn}.
For $\overline{\mathbb Z}^s_2(k)$, the predicted $\cos \delta_D$ (\ref{eq:G-relation-b}) is
larger than (\ref{eq:G-relation-a}) by a factor of $2$. Consequently, the peak moves to
around $\pm 123^\circ$ ($\pm 57^\circ$).

The distribution of the leptonic Jarlskog invariant $J_\nu$ is shown in Fig.~\ref{fig:DJ-J}.
These predictions can be tested at T2K \cite{t2k} and at NO$\nu$A \cite{nova}. 

{\it Conclusions} --
Phenomenological consequences of the residual $\mathbb Z^s_2(k)$ and $\overline{\mathbb Z}^s_2(k)$ 
symmetries are compared with data and global fits. Although not independent, their predictions are
different. A large reactor angle $\theta_x$ peaking around $3^\circ$ or $6^\circ$ which is consistent with 
T2K, MINOS and Double Chooz can be obtained and the Dirac {\tt CP} phase 
$\delta_D$ has peaks at $\pm 74^\circ$ ($\pm 106^\circ$) or $\pm 123^\circ$ ($\pm 57^\circ$) 
in excellent agreement with the latest global fits. 
This provides the first strong and direct support for 
$\mathbb Z^s_2(k)$ and $\overline{\mathbb Z}^s_2(k)$ as residual symmetries of neutrino mixing.
Further confirmation may come from the measurement of the
leptonic Jarlskog invariant $J_\nu$ at T2K or NO$\nu$A.

{\it Acknowledgments} --
It is our pleasure to thank Karol Lang for discussions about MINOS, Wade Fisher for 
discussions about handling asymmetric errors and Jim Linnemann for help with extracting the $\sin \theta_x$
distributions.
Also we greatly appreciate correspondence with E. Lisi, M. Maltoni, T. Schwetz, and  J.W.F. Valle
concerning the fits to the data.
DAD was supported in part by the U. S. Department of Energy under grant No. DE-FG03-93ER40757
and WWR was supported in part by the National Science Foundation under Grant PHY-1068020.

\end{document}